\documentclass[aps,english,showpacs,twocolumn,prd]{revtex4-1}
\usepackage{amsfonts}
\usepackage{amssymb}
\usepackage{amsmath}
\usepackage{tikz}
\usetikzlibrary[decorations.pathmorphing]
\usepackage{graphicx}
\usepackage{epsfig}
  \usepackage{filecontents}
\begin{document}

\title{Detecting the processes of colliding plane gravitational waves by electromagnetic response signals}
\author{Dong-dong Wei$^a$}
\email{wdd@mail.nankai.edu.cn}
\author{Xin-he Meng$^a$}
\email{xhm@nankai.edu.cn}
\author{Bin Wang$^a$}
\email{wb@mail.nankai.edu.cn}
\affiliation{$^a$~School of Physics, Nankai University, Tianjin 300071, China}
\begin{abstract}
In this paper, we have considered how to detect the processes of plane gravitational waves colliding. The degenerate Ferrari-Ibanez solution describes the collision of plane gravitational waves with aligned linear polarization, and the solution of the interaction region is Schwarzschild-like metric by taking coordinate transformation, which impels us more interesting to detect the process to explore. We have calculated explicitly out the solutions of the electromagnetic field produced by the plane gravitational wave and the colliding region of plane gravitational waves perturbing a weak magnetic field background. The magnitudes are so small that the likelihood of detecting gravitational waves is only just emergence within the range of the most high-level modern apparatus such as the further upgraded aLIGO, but we can judge whether the collision process has occurred or not by measuring the amplitude and waveform of electromagnetic wave properties. Moreover, detecting electromagnetic waves can offer a new method to verify general relativity.
\end{abstract}

\maketitle

\section{Introduction}
As it is well known that the gravitation waves from binary star coalescence have been repeatedly detected \cite{Abbott:2016blz}\cite{Abbott:2016nmj}\cite{Abbott:2017vtc}\cite{Abbott:2017oio}\cite{Abbott:2017gyy}, especially the associated electromagnetic signals from binary neutron star mergers \cite{TheLIGOScientific:2017qsa}, which inspires us to wonder if we can use relatively easily located corresponding electromagnetic signals to reveal the weak gravitation wave properties. The O3 phase of aLlGO and aVIRGO is producing more opportunities to explore gravity research boundaries.

One of the central study fields in general relativity is the collision of gravitational plane waves. As we know, gravitational waves(GWs) described by Einstein’s equation are highly non-linear, which is bound to pass through each other having a significant interaction. Of course, detecting the interaction region is also the best way to verify the Einstein theory. 
In 1971, Penrose and Khan have firstly found out an exact solution of the interaction of two impulsive gravitational waves\cite{Khan:1971vh}, and in the same years, Szekeres has derived the equations governing the collision of two plane gravitational waves using Newman-Penrose techniques\cite{szekeres1972colliding}. In 1986, Ferrari has constructed a class of soliton solution on colliding plane gravitational waves (CPWs)\cite{Ferrari1987On}. This class of solution in vacuum Einstein equations describe a space-time of two plane gravitational waves colliding, and its degenerate case is locally isometric to the Schwarzschild solution which impels us to study it further. In 1987, Chandrasekhar, Subrahmanyan and Xanthopoulos\cite{Chandrasekhar:1987rt} have demonstrated that the coupling of the gravitational waves to EM waves does not affect the development of a horizon nor a time-like curvature singularity alone hyperbolic arcs. Moreover, Miquel Dorca and Enric Verdaguer have considered quantum field interacting with colliding pane waves\cite{dorca1993quantum}\cite{Dorca:1994pf}\cite{dorca1997quantum}.

Existing large scale magnetic fields in the universe\cite{Marklund:2000zs} \cite{melvin1965dynamics} \cite{bonnor1954static} \cite{Barrabes:2010tr} \cite{Wen:2014wxa} \cite{Bamba:2018cup} have provided us an opportunity to discuss the CPWs with magnetic fields. Marklund, Dunsby, and Brodin \cite{Marklund:2000zs} have shown that the gravitational waves propagating in the spacetime with a weak magnetic field can produce electromagnetic waves. This work is to concern about how to detect the colliding process really happening by using the electromagnetic signal. We choosing the soliton solution of  Ferrari and Ibanez have three reasons. Firstly, we like to consider the PGWs rather than spherical and cylindrical waves, because the PGWs have a Killing vector which is represented for translational symmetry. 
Secondly, the soliton solution ensures the energy of PGWs conserving in the propagating direction while guaranteeing we do not measure zero electromagnetic(EM) waves produced by the PGWs perturbing the magnetic field to a long distant observer\cite{Griffiths:1991zp}. Thirdly,it is rather easy to calculate. Therefore, we consider a pair of PGWs coming from the infinity and colliding with each other. But in the region of PGWs, we assume that there exist a weak magnetic field.     

 This paper is organized as follows. After a brief introduction of degenerate Ferrari-Ibanez solutions, in section \uppercase\expandafter{\romannumeral3}, we work out solutions of the perturbed EM field in the colliding region. In section \uppercase\expandafter{\romannumeral4}, we also give the solutions of the disturbed EM field in the region of PGWs. In the last section, the summary and conclusion are given.

\section{The soliton solution of Ferrari and Ibanez}

The general solution for the Ferrari and Ibanez can be expressed in the metric\cite{Ferrari1987On}\cite{V1978Integration}:

  \begin{eqnarray}
  \label{1}
{{\it ds}}^{2}=&-4\,{\frac { \left( 1+u+v \right) ^{2\,{s}^{2}} \left( 
1-v-u \right) ^{2\,{q}^{2}}{\it du}\,{\it dv}}{\sqrt {-{u}^{2}-{v}^{2}
+1}}}\nonumber
\\&+{\frac { \left( -{u}^{2}-{v}^{2}+1 \right) ^{2\,q} \left( 1-v-u
 \right) {{\it dx}}^{2}}{1+u+v}}\nonumber \\&+{\frac { \left( -{u}^{2}-{v}^{2}+1
 \right) ^{2\,s} \left( 1+u+v \right) {{\it dy}}^{2}}{1-v-u}}
  \end{eqnarray}
where $s+q=1$

This metric describes the collision of plane gravitational waves with aligned linear polarization. Specially, when we take $s=q=\frac{1}{2}$, the metric (\ref{1})  becomes the Khan-Penrose solution\cite{Khan:1971vh} in the colliding region between two impulsive plane gravitational waves. Solution (\ref{1}) can follow Khan-Penrose divide the space into four regions, taking $u\to u \text{H(u)} $ , $v\to v \text{H(v)}$ where H(x) means the Heaviside step function. Then the solution can be extended into region I, II, III(see Fig.1). The region II and III describe the spacetime of a PGW while the region I portray the background spacetime. Finally, the metric of three regions can be expressed :

\begin{eqnarray}
I: ~\ ~\ {{\it ds}}^{2}=-4\,{\it du}\,{\it dv}+{{\it dx}}^{2}+{{\it dy}}^{2}
\end{eqnarray}

\begin{eqnarray}
II: ~\ ~\ {{\it ds}}^{2}=-4\,{\frac { \left( 1+v \right) ^{2\,{s}^{2}} \left( 1-
v \right) ^{2\,{q}^{2}}{\it du}\,{\it dv}}{\sqrt {-{v}^{2}+1}}}\nonumber \\+{
\frac { \left( -{v}^{2}+1 \right) ^{2\,q} \left( 1-v \right) {{\it dx}
}^{2}}{1+v}}\nonumber \\+{\frac { \left( -{v}^{2}+1 \right) ^{2\,s} \left( 1+v
 \right) {{\it dy}}^{2}}{1-v}}
\end{eqnarray}

\begin{eqnarray}
III: ~\ ~\ {{\it ds}}^{2}=-4\,{\frac { \left( 1+u \right) ^{2\,{s}^{2}} \left( 1-
u \right) ^{2\,{q}^{2}}{\it du}\,{\it dv}}{\sqrt {-{u}^{2}+1}}}\nonumber \\+{
\frac { \left( -{u}^{2}+1 \right) ^{2\,q} \left( 1-u \right) {{\it dx}
}^{2}}{1+u}}\nonumber \\+{\frac { \left( -{u}^{2}+1 \right) ^{2\,s} \left( 1+u
 \right) {{\it dy}}^{2}}{1-u}}
\end{eqnarray}

In the solution of Ferrari and Ibanez, we only study the degenerate case which $s=1,q=0 ~\ or ~\ s=0,q=1 $. Because the degenerate case can be avoided the Cauchy horizon in the colliding region. For example , the structure of Khan-Penrose have a Cauchy horizon\cite{Griffiths:1991zp}. If we consider the EM waves created by the perturbed magnetic field in the neighborhood of the Cauchy horizon, the EM waves will suffer a mass inflation or the spacetime will become instability\cite{Yurtsever:1987gv}\cite{poisson1989inner}\cite{Yurtsever:1988vc}. The mass inflation and the instability of the spacetime definitely will change the metric (\ref{1}), making the process complicated. For the above reasons, we only consider the degenerate case which is given by (\ref{1}).Taking the coordinate transformation:
\begin{eqnarray}
& u=1/2\,\cos \left( \pi-r-t \right) \\
& v=1/2\,\sin \left( t-r \right) 
\end{eqnarray}

The degenerate metric can be expressed as :
\begin{eqnarray}
\label{7}
IV:&{ds}^2=- \frac{(\sin(t)+1)^2 }{2}(dt^2-dr^2)\nonumber \\&+\frac{1-\sin(t)}{1+\sin (t)}{dx}^2+({\sin(t)}+1)^2 \cos^2 (r) {dy}^2
\end{eqnarray}

And the region II and III also can be expressed as:
\begin{eqnarray}
\label{8}
II: ~\ ~\ {ds}^2=- \frac{(\sin(t+r)+1)^2 }{2}(dt^2-dr^2)\nonumber \\+\frac{1-\sin(t+r)}{1+\sin (t+r)}{dx}^2\nonumber \\+ ({\sin(t+r)}+1)^2 \cos^2 (t+r) {dy}^2
\end{eqnarray} 
\begin{eqnarray}
\label{9}
III: ~\ ~\ {ds}^2=- \frac{(\sin(t-r)+1)^2 }{2}(dt^2-dr^2)\nonumber\\+\frac{1-\sin(t-r)}{1+\sin (t-r)}{dx}^2\nonumber \\+  ({\sin(t-r)}+1)^2 \cos^2 (t-r) {dy}^2
\end{eqnarray}

 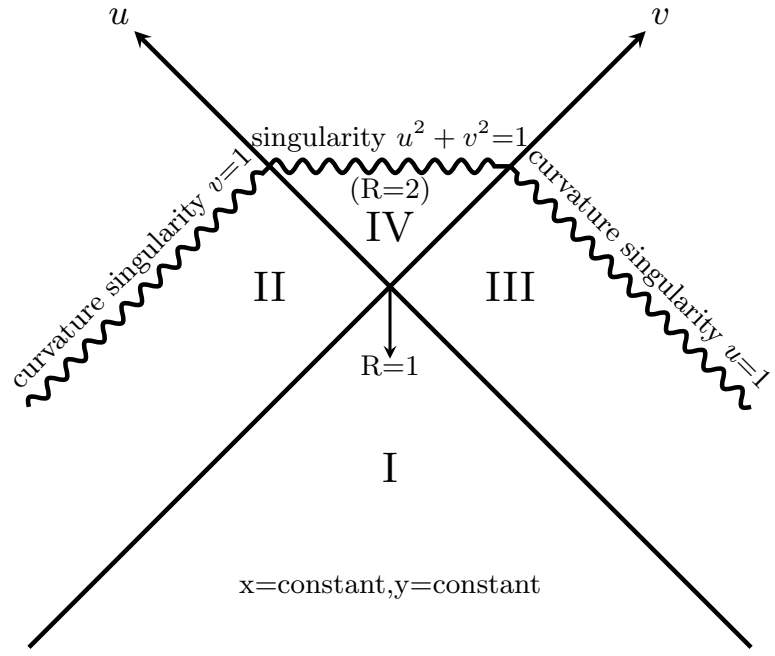
\begin{figure}
 \label{img}
\begin{tikzpicture}[scale=0.8]
 \draw [thick] [line width=0.6mm] [-stealth] (-6,-6) -- (4.25,4.25);
\draw [thick] [line width=0.6mm] [-stealth] (6,-6) -- (-4.25,4.25);
\draw [decorate,decoration={coil,aspect=0}] [line width=0.6mm] (-6,-2) -- node[sloped,above,scale=1.2] {curvature singularity $v$=1}(-2,2);
\draw [decorate,decoration={coil,aspect=0}] [line width=0.6mm] (6,-2) -- node[sloped,above,scale=1.2] {curvature singularity $u$=1} (2,2);
\draw [decorate,decoration={coil,aspect=0}] [line width=0.6mm] (-2,2) -- (2,2);
\draw [thick] [line width=0.4mm] [-stealth] (0,0) -- (0,-1.2);
\node at (0,-1.3)[scale=1.2] {R=1};
\node at (0,1.6) [scale=1.2] {(R=2)};
\node at (0,-3) [scale=1.8] {I};
\node at (-2,0) [scale=1.8] {II};
\node at (0,1) [scale=1.8] {IV};
\node at (2,0) [scale=1.8] {III};
\node at (0,2.5) [scale=1.2] {singularity $u^2+v^2$=1};
\node at (4.5,4.5) [scale=1.5] {$v$};
\node at (-4.5,4.5) [scale=1.5] {$u$};
\node at (0,-5) [scale=1.2] {x=constant,y=constant};
\end{tikzpicture}
\vspace*{3mm} 
 \caption{The spacetime of CPWs is separated into four valid regions.The PGWs collide at the $R=M$ or $u=v=0$ and the colliding region reaches its maximum value at $R=2M$ or $u^2+v^2=1$ } 

\end{figure}

As is shown in Fig.2, it is easier to understand the colliding process with a magnetic field background in the spatial picture.

 \begin{figure}
  \centering
  \includegraphics[width=.4\textwidth,trim=50 200 100 200,clip]{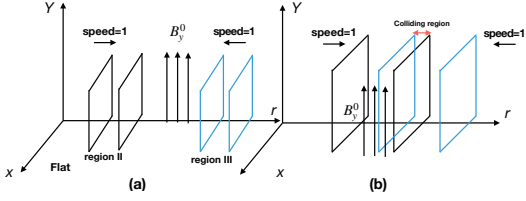} 
  \caption{(a) portrays two PGWs propagating in the opposite way with a magnetic field background. On the left side of the PGW, we call it as region II. On the right side of the PGW, we call it as region III. And the background in the process is called region I. (b) depicts that the process of CPWs whose colliding region is called region IV.  } 
  \label{img2} 
\end{figure}

\section{Solution of EM waves in the colliding region}

In this part, we will discuss the situation that the process of CPWs has happened at a magnetic field region in the $y$ axis. The magnetic field will be perturbed by the PGW as well as colliding region, creating weak EM wave signals. Thus, the EM energy-momentum tensor can be written as:   
\begin{eqnarray}
\label{eq1}
&~&F_{\alpha \beta } ={F^{_{(0)}}_{\alpha\beta}}+ {F^{_{(1)}}_{\alpha\beta}}\nonumber\\
~\nonumber\\
&=&\left( {{\begin{array}{*{20}c}
 0  &  {E^{_{(1)}}_r}  &{E^{_{(1)}}_x} & &{E^{_{(1)}}_y}&  \\
 ~\nonumber\\
-{E^{_{(1)}}_r} & 0  &  (-{B^{_{(0)}}_{y}}-{B^{_{(1)}}_y}) & &{B^{_{(1)}}_x} & \\
~\nonumber\\
-{E^{_{(1)}}_x}  & ({B^{_{(0)}}_{y}}+{B^{_{(1)}}_y})  & 0  & -&{B^{_{(1)}}_r}&  \\
~\nonumber\\
-{E^{_{(1)}}_y} & -{B^{_{(1)}}_x}   &  {B^{_{(1)}}_r}    &  &0& \\
\end{array} }} \right)\nonumber\\
\end{eqnarray}
~\\
and the Maxwell equations in the curved spacetime :
\begin{eqnarray}
\label{eq2}
&\nabla_{\nu}F^{\mu\nu}=-4\pi J^{\mu},\nonumber\\
~\nonumber\\
&\nabla_{[\alpha}F_{\mu\nu]}= 0
\end{eqnarray}
~\\
In this process $J^{\mu}$ is zero.\\
We will take Eqs. (\ref{7}-\ref{9}) into the Maxwell equations, but this process will generate complicated equations which is hard to express in this paper. So we rewrite the metrics (\ref{7}-\ref{9}) into a simple form:
\begin{eqnarray}
\label{new}
 ds^2=-\sigma(t,r) dt^{2}+\sigma(t,r) dr^{2}\nonumber \\+\gamma(t,r) dx^{2}+\psi(t,r) dy^{2}
\end{eqnarray}
where the $\sigma, ~\ \gamma, ~\ \psi$ denote $g_{0 0}$ or $g_{1 1}$, $g_{2 2} ~\ and ~\ g_{3 3} $ respectively in the metrics (\ref{7}-\ref{9}).
~\\
By expanding the Maxwell equations, we have eight equations:
\begin{eqnarray}
\label{13}
{\it E^{(1)}_{r}} \left( t,r \right)  \left( {\frac {\partial }{\partial r}}
\psi \left( t,r \right)  \right) \sigma \left( t,r \right) \gamma
 \left( t,r \right) ~\nonumber \\ +2\, \left( {\frac {\partial }{\partial r}}{\it 
E^{(1)}_{r}} \left( t,r \right)  \right) \sigma \left( t,r \right) \gamma
 \left( t,r \right) \psi \left( t,r \right)~\nonumber \\ 
+ {\it E^{(1)}_{r}} \left( t,r
 \right)  \left( {\frac {\partial }{\partial r}}\gamma \left( t,r
 \right)  \right) \sigma \left( t,r \right) \psi \left( t,r \right) ~\nonumber \\ -2
\,{\it E^{(1)}_{r}} \left( t,r \right) \gamma \left( t,r \right)  \left( {
\frac {\partial }{\partial r}}\sigma \left( t,r \right)  \right) \psi
 \left( t,r \right) =0
\end{eqnarray}
\begin{eqnarray}
\label{14}
 {\it E^{(1)}_{r}} \left( t,r \right)  \left( {\frac {\partial }{\partial t}}
\psi \left( t,r \right)  \right) \sigma \left( t,r \right) \gamma
 \left( t,r \right) ~\nonumber \\+{\it E^{(1)}_{r}} \left( t,r \right)  \left( {\frac {
\partial }{\partial t}}\gamma \left( t,r \right)  \right) \sigma
 \left( t,r \right) \psi \left( t,r \right) ~\nonumber \\+2\,\sigma \left( t,r
 \right)  \left( {\frac {\partial }{\partial t}}{\it E^{(1)}_{r}} \left( t,r
 \right)  \right) \gamma \left( t,r \right) \psi \left( t,r \right) ~\nonumber \\-2
\,{\it E^{(1)}_{r}} \left( t,r \right)  \left( {\frac {\partial }{\partial t}}
\sigma \left( t,r \right)  \right) \gamma \left( t,r \right) \psi
 \left( t,r \right) =0
\end{eqnarray}
\begin{eqnarray}
\label{15}
&{\it B^{(0)}_{y}} \left( {\frac {\partial }{\partial r}}\psi \left( t,r
 \right)  \right) \gamma \left( t,r \right) \nonumber \\&-{\it B^{(0)}_{y}} \left( {
\frac {\partial }{\partial r}}\gamma \left( t,r \right)  \right) \psi
 \left( t,r \right) \nonumber \\& + \left( {\frac {\partial }{\partial r}}\psi
 \left( t,r \right)  \right) \gamma \left( t,r \right) {\it B^{(1)}_{y}}
 \left( t,r \right) \nonumber \\&+2\left( {\frac {\partial }{\partial r}}{\it 
B^{(1)}_{y}} \left( t,r \right)  \right) \gamma \left( t,r \right) \psi
 \left( t,r \right)\nonumber \\ &- \left( {\frac {\partial }{\partial r}}\gamma
 \left( t,r \right)  \right) {\it B^{(1)}_{y}} \left( t,r \right) \psi \left( 
t,r \right) +\gamma \left( t,r \right) {\it E^{(1)}_{x}} \left( t,r \right) {
\frac {\partial }{\partial t}}\psi \left( t,r \right) \nonumber \\&+2\gamma
 \left( t,r \right)  \left( {\frac {\partial }{\partial t}}{\it E^{(1)}_{x}}
 \left( t,r \right)  \right) \psi \left( t,r \right) \nonumber \\&-{\it E^{(1)}_{x}}
 \left( t,r \right)  \left( {\frac {\partial }{\partial t}}\gamma
 \left( t,r \right)  \right) \psi \left( t,r \right) =0&
\end{eqnarray}

\begin{eqnarray}
\label{16}
  &-\left( {\frac {\partial }{\partial r}}\psi \left( t,r \right) 
 \right) \gamma \left( t,r \right) {\it B^{(1)}_{x}} \left( t,r \right) \nonumber \\&+2\,
 \left( {\frac {\partial }{\partial r}}{\it B^{(1)}_{x}} \left( t,r \right) 
 \right) \gamma \left( t,r \right) \psi \left( t,r \right) \nonumber \\&+ \left( {
\frac {\partial }{\partial r}}\gamma \left( t,r \right)  \right) \psi
 \left( t,r \right) {\it B^{(1)}_{x}} \left( t,r \right) -2\,\gamma \left( t,r
 \right) \psi \left( t,r \right) {\frac {\partial }{\partial t}}{\it 
E^{(1)}_{y}} \left( t,r \right) \nonumber \\&+\gamma \left( t,r \right) {\it E^{(1)}_{y}} \left( t,
r \right) {\frac {\partial }{\partial t}}\psi \left( t,r \right) -\nonumber \\&
 \left( {\frac {\partial }{\partial t}}\gamma \left( t,r \right) 
 \right) \psi \left( t,r \right) {\it E^{(1)}_{y}} \left( t,r \right) =0&
\end{eqnarray}
\begin{eqnarray}
\frac{\partial {B^{_{(1)}}_r} }{\partial r}=0
\end{eqnarray}
\begin{eqnarray}
\frac{\partial {B^{_{(1)}}_r} }{\partial t}=0
\end{eqnarray}
\begin{eqnarray}
\frac{\partial{E^{_{(1)}}_y}}{\partial r}=\frac{\partial{B^{_{(1)}}_x}}{\partial t}
\end{eqnarray}
\begin{eqnarray}
\label{20}
\frac{\partial{E^{_{(1)}}_x}}{\partial r}=-\frac{\partial{B^{_{(1)}}_y}}{\partial t}
\end{eqnarray}
and we have omitted second and higher order infinitesimal terms, it gives:
\begin{equation}
\label{EE}
{\frac {\partial ^{2}}{\partial {t}^{2}}}{\it E^{(1)}_{x}} \left( t,r
 \right)-{\frac {\partial ^{2}}{\partial {r}^{2}}}{\it E^{(1)}_{x}} \left( t,r \right)  =-\frac{1}{2}\,{\it B^{(0)}_{y}}\, \left( {\frac {\psi_{r}}{\psi}}-{\frac {\gamma_{r}}{
\gamma}} \right) _{t}
\end{equation}
\begin{equation}
\label{BB}
{\frac {\partial ^{2}}{\partial {t}^{2}}}{\it B^{(1)}_{y}} \left( t,r
 \right)-{\frac {\partial ^{2}}{\partial {r}^{2}}}{\it B^{(1)}_{y}} \left( t,r \right)  =\frac{1}{2}\,{\it B^{(0)}_{y}}\, \left( {\frac {\psi_{r}}{\psi}}-{\frac {\gamma_{r}}{
\gamma}} \right) _{r}
\end{equation}
~\\
where the subscript $t$ and $r$ in the right-hand side of the above equations is denoted as ordinary derivatives with t and r respectively. 

One may concerned about that the second and higher order may not neglect, since there is a curvature sigularity, which making these electromagnetic have divergences. However,this method can be use in out of the neighborhood of curvature singularity and it is enough to analyst the colliding process.

Combining the Eq. (\ref{15}) and the initial condition, we have:
\begin{eqnarray}
\label{partialEquBound}
{E^{_{(1)}}_x}|_{_{t=0}}&=&0,~~\frac{\partial{E^{_{(1)}}_x}}{\partial t}|_{_{t=0}}=-\frac{1}{2}\, \left( {\frac {\psi_{r}}{\psi}}-{\frac {\gamma_{r}}{
\gamma}} \right) \big|_{_{t=0}}{B_{y}^{_{(0)}}} , \nonumber\\
~\nonumber\\
{B^{_{(1)}}_y}|_{_{t=0}}&=&0, ~~\frac{\partial{B^{_{(1)}}_y}}{\partial t}|_{_{t=0}}=0
\end{eqnarray}
~\\

 Eqs. (\ref{13}, \ref{14} and \ref{16}) combine with the initial condition only have zero results.

By using analytical way, the solutions of (\ref{EE} and \ref{BB}) can be expressed as :
\begin{eqnarray}
\label{E}
{E^{_{(1)}}_x}&=&\frac{1}{2}\int_{x-t}^{x+t}H(\xi)d\xi \nonumber\\
~\nonumber\\
&+&\frac{1}{2}\int_{0}^{t}\int_{x-(t-\tau)}^{x+(t-\tau)}F(\xi, \tau)d\xi d\tau,\\
~\nonumber\\
~\nonumber\\ \label{B}
{B^{_{(1)}}_y}&=& \frac{1}{2}\int_{0}^{t}\int_{x-(t-\tau)}^{x+(t-\tau)}G(\xi, \tau)d\xi d\tau,
\end{eqnarray}
\qquad  \qquad  where,
\begin{eqnarray}
\label{H}
&~&H(\xi)=-\frac{1}{2}{\it B^{(0)}_{y}}\, \left( {\frac {\psi_{\xi}}{\psi}}-{\frac {\gamma_{\xi}}{
\gamma}} \right)_{\xi}\big|_{_{t=0}} ~\nonumber\\
&~&F(\xi, \tau)=-\frac{1}{2}\,{\it B^{(0)}_{y}}\, \left( {\frac {\psi_{\xi}}{\psi}}-{\frac {\gamma_{\xi}}{
\gamma}} \right)_{\xi \tau} \nonumber\\
~\nonumber\\
&~&G(\xi, \tau)= \frac{1}{2}{\it B^{(0)}_{y}}\, \left( {\frac {\psi_{\xi}}{\psi}}-{\frac {\gamma_{\xi}}{
\gamma}} \right)_{\xi \xi}  
\end{eqnarray}
~\\
the spacetime of CPWs can be separated into three regions. First, we discuss the colliding region($0 \le u\le 1,~\ 0 \le v \le1$).\\
Putting $r \to \xi, ~\ ~\ t\to \tau$, in the colliding region we have:\\
\begin{eqnarray}
{\frac {\psi_{\xi}}{\psi}}-{\frac {\gamma_{\xi}}{
\gamma}}=-2\,{\frac {\sin \left( \xi \right) }{\cos \left( \xi \right) }}
\end{eqnarray}
And put it into the Eqs. (\ref{E}) and (\ref{B}), we can easily get that(which have been checked by Maple 2017 PDEtools):\\
\begin{eqnarray}
\label{28}
{E^{_{(1)}}_x}=\frac{{B^{_{(0)}}_y}}{2} ( \ln (\sin (r-t))- \ln (\sin (r+t))\nonumber \\+\ln \left(-\cot (r-t)\right)-\ln \left(-\cot (r+t)\right)) 
\end{eqnarray}
\begin{eqnarray}
\label{29}
&{B^{_{(1)}}_y}=\frac{1}{2}\, {B^{_{(0)}}_y} ( \ln  \left( \cos \left( r-t
 \right)  \right) \nonumber \\&+\ln  \left( \cos \left( t+r \right)  \right) -2\,
\ln  \left( \cos \left( r \right)  \right)) 
\end{eqnarray}
We have noticed the EM field does not exist singularity at $t=\frac{\pi}{2}$ (see Fig.3). This phenomenon can be explained that there is no curvature singularity at $t=\frac{\pi}{2}$. Of course, it is properly because of  $t=\frac{\pi}{2}$ corresponding to the event horizon of the Schwarzschild black hole which does not curvature singularity \cite{Ferrari1987On}.   
    \begin{figure}
  \centering
  \label{3}
 \includegraphics[width=.3\textwidth]{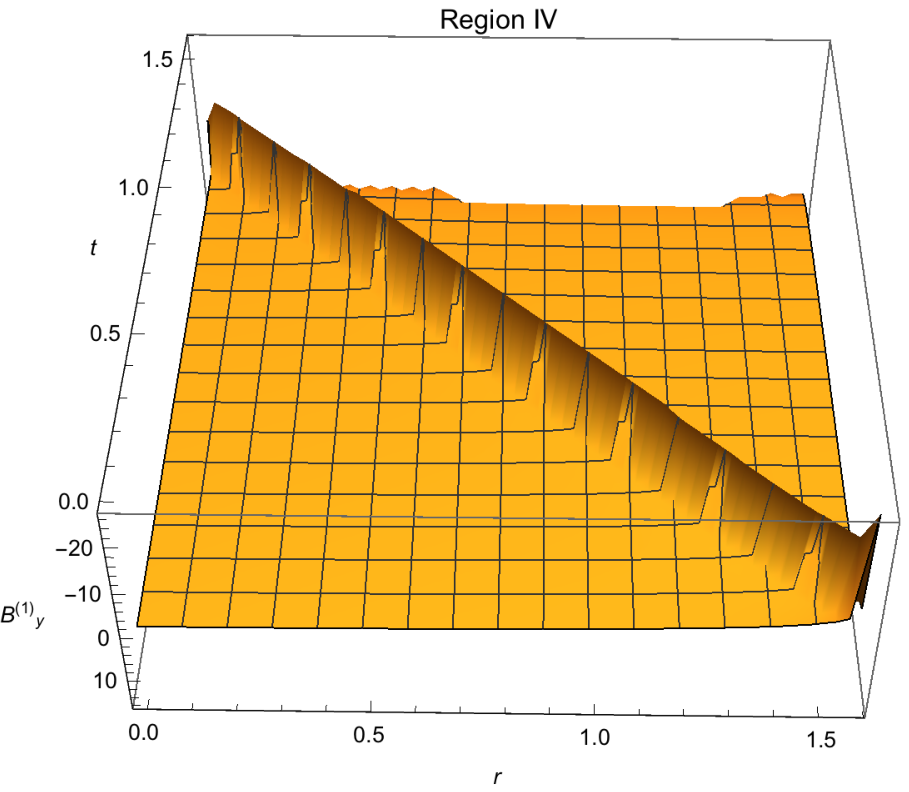} 
 \includegraphics[width=.3\textwidth]{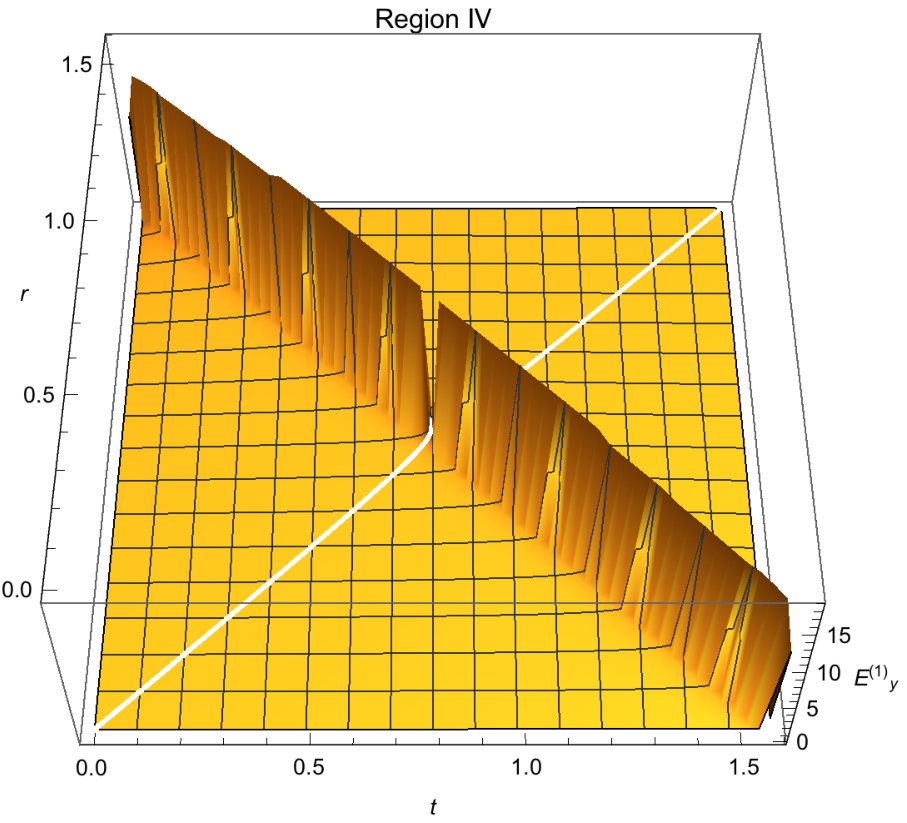}
    \caption{The picture above describes the Eq. (\ref{28}), and the figure below describes the Eq. (\ref{29})  }
\end{figure}

\begin{figure}
  \centering
  \includegraphics[width=.3\textwidth]{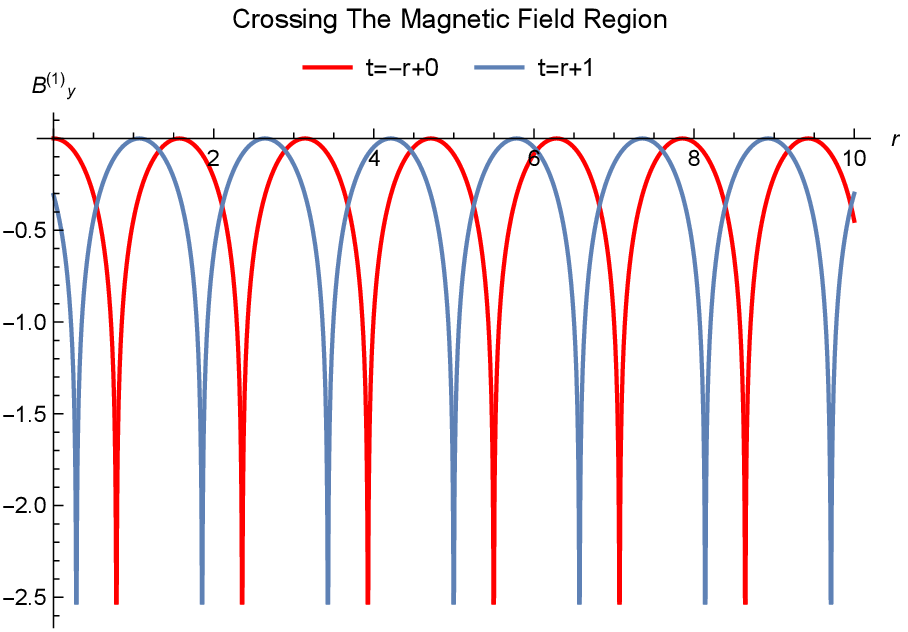} 
    \includegraphics[width=.3\textwidth]{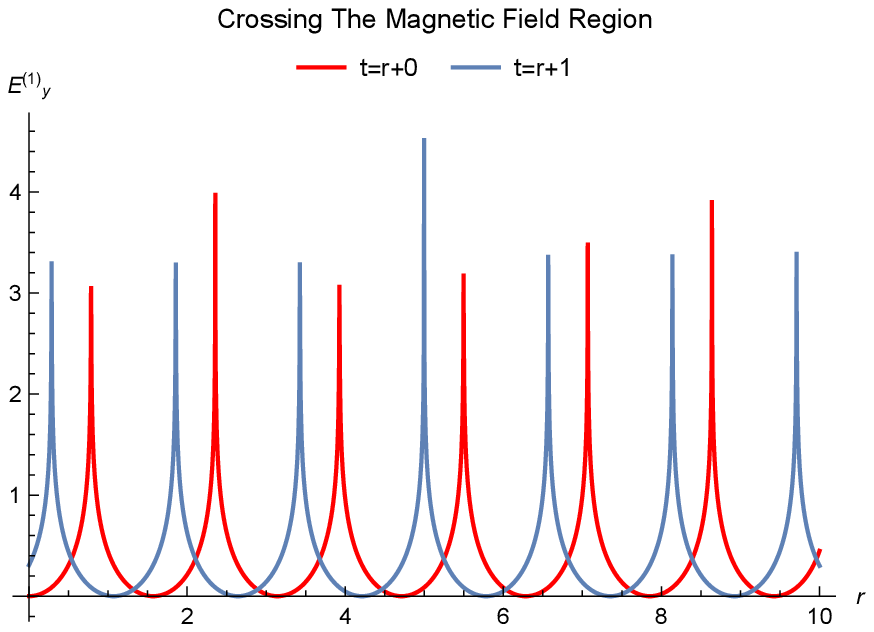} 
    \caption{ Diagrams describe the EM waves propagating in the u direction with the magnetic field background, and the above and below picture depict the magnetic field and electric field respectively      }
\end{figure}
\begin{figure}
  \centering
  \includegraphics[width=.3\textwidth]{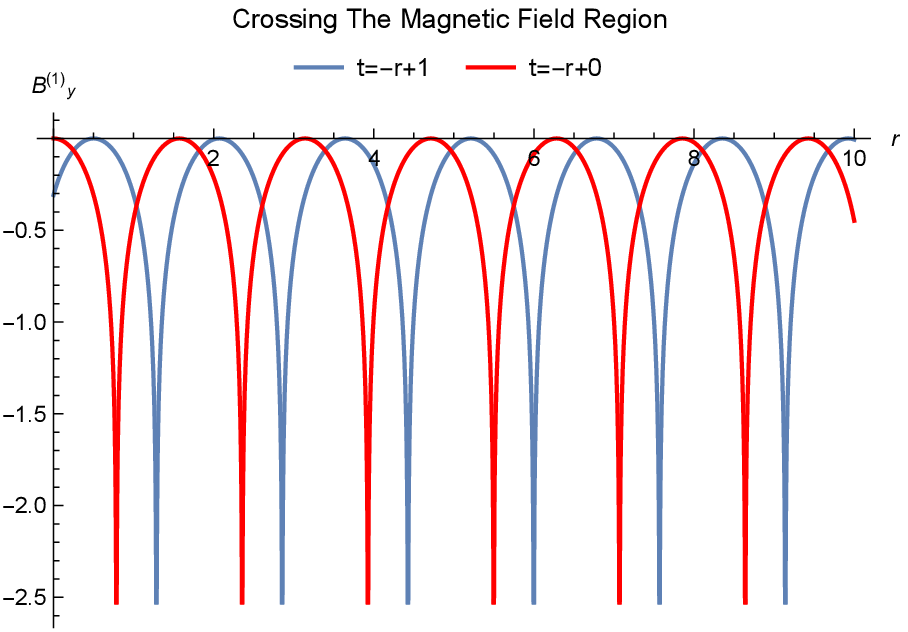} 
    \includegraphics[width=.3\textwidth]{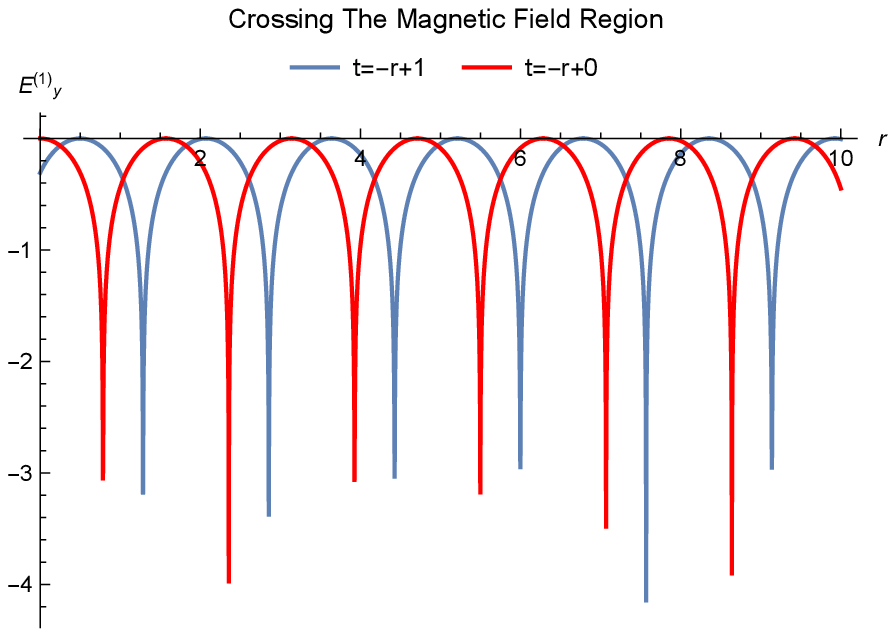} 
    \caption{Diagrams describe the EM waves propagating in the v direction with the magnetic field background, and the above and below picture depict the magnetic field and electric field respectively   }
\end{figure}

Because the coordinate $(t,r,x,y)$ can not describes the entire spacetime, the global structure of the spacetime has been determined by Hayward\cite{Hayward:1989sq}. Putting $t=\arcsin (R-1),~\ T=\sqrt{2}x,~\ \theta=\frac{\pi}{2}-r, ~\  \phi=\sqrt{2}y $, the metric can be recognized as the Schwarzschild solution. The extended solution is more larger than before. We also put this coordinate transformation into the Eq.(\ref{28} and \ref{29}), which can let us observe the extension of EM waves clearly. When R=1 to R=2, the EM waves propagate in $\theta$ direction(as you can see in the Eqs.(\ref{EE} and \ref{BB}). However, when $R> 2$, the EM waves which extended by the coordinate transformation can not propagate in the $\theta$ direction. Because $\frac {\partial } {\partial R }$ is a spacelike Killing vector which causes the Maxwell Eqs. (\ref{EE} and \ref{BB})  no more propagation equations. 
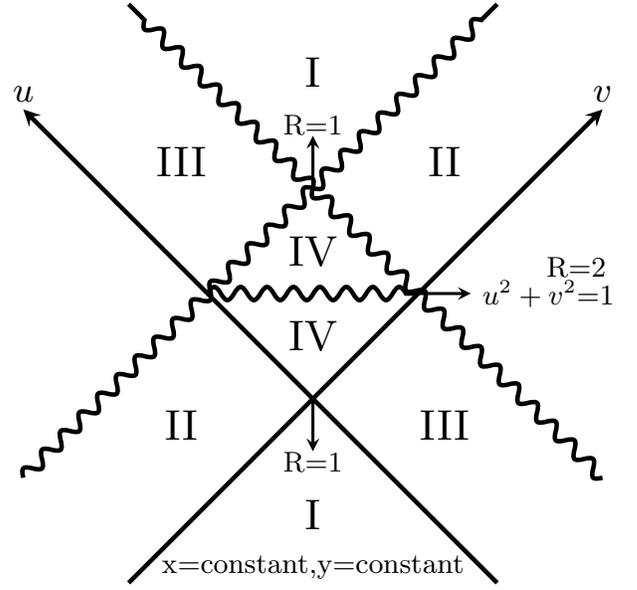
\begin{figure}
\begin{tikzpicture}[scale=0.7]
\draw [thick] [line width=0.6mm] [-stealth] (-3.5,-5.5) -- (5.5,3.5);
\draw [thick] [line width=0.6mm] [-stealth] (3.5,-5.5) -- (-5.5,3.5);
\draw [decorate,decoration={coil,aspect=0}] [line width=0.6mm] (-5.5,-3.5) -- (3.5,5.5);
\draw [decorate,decoration={coil,aspect=0}] [line width=0.6mm] (5.5,-3.5) -- (-3.5,5.5);
\draw [decorate,decoration={coil,aspect=0}] [line width=0.6mm] (-2,0) -- (2,0);
\draw [thick] [line width=0.4mm] [-stealth] (2,0) -- (3,0)node[right,scale=1.2] { $u^2+v^2$=1};
\draw [thick] [line width=0.4mm] [-stealth] (0,-2) -- (0,-3);
\draw [thick] [line width=0.4mm] [-stealth] (0,2) -- (0,3);
\node at (0,3.2)[scale=1.2] {R=1};
\node at (5.5,3.8) [scale=1.5] {$v$};
\node at (-5.5,3.8) [scale=1.5] {$u$};
\node at (0,-3.2)[scale=1.2] {R=1};
\node at (5,0.5)[scale=1.2] {R=2};
\node at (0,-4.2) [scale=1.8] {I};
\node at (-2.5,-2.5) [scale=1.8] {II};
\node at (0,-0.8) [scale=1.8] {IV};
\node at (2.5,-2.5) [scale=1.8] {III};
\node at (0,4.2) [scale=1.8] {I};
\node at (-2.5,2.5) [scale=1.8] {III};
\node at (0,0.8) [scale=1.8] {IV};
\node at (2.5,2.5) [scale=1.8] {II};
\node at (0,-5.2) [scale=1.2] {x=constant,y=constant};
\end{tikzpicture}
\vspace*{3mm}    
	\caption{The sketching is showed us the time symmetry extension.}
\end{figure}

As suggested by the Dong-dong Wei\cite{weidongdong}, the lower energy scale case and the high energy scale case should correspond to the the Schwarzschild exterior solution and the time symmetry solution respectively. When we take the Schwarzschild exterior solution, EM waves can not propagate in the $R\geq 2$ and the EM waves propagate in the $0<R<2$ can not be detected. Hence, the lower energy scale case may not detectable by this way. Another situation is taken the symmetric solution. As we can see the EM waves described by the Eqs. (\ref{28} and \ref{29}) will propagate in u and v direction by taking time symmetry. Futhermore, the region II, III and IV interacting with a magnetic field is finite. When the PGWs coupling with the EM waves totally cross the magnetic field region, the Eqs. (\ref{EE} and \ref{BB}) have become:
\begin{equation}
\label{EE11}
{\frac {\partial ^{2}}{\partial {t}^{2}}}{\it E^{(1)}_{x}} \left( t,r
 \right)-{\frac {\partial ^{2}}{\partial {r}^{2}}}{\it E^{(1)}_{x}} \left( t,r \right)  =0
\end{equation}
\begin{equation}
\label{BB11}
{\frac {\partial ^{2}}{\partial {t}^{2}}}{\it B^{(1)}_{y}} \left( t,r
 \right)-{\frac {\partial ^{2}}{\partial {r}^{2}}}{\it B^{(1)}_{y}} \left( t,r \right)  =0
\end{equation}
~\\
 In order to satisfy these equations, we consider that:
 \begin{eqnarray}
 \label{ac}
E'^{(1)}_{x}&=&\frac{B'^{(0)}_{y}}{4} \left(2 \ln (\sin (r-t))+\ln \left(-\cot^2(r-t)\right)\right) \nonumber \\
B'^{(1)}_{y}&=&\frac{B'^{(0)}_{y}}{2} \ln (\cos (r-t))
\end{eqnarray}
~\
 \begin{eqnarray}
 \label{ab}
E''^{(1)}_{x}&=&\frac{B'^{(0)}_{y}}{4} \left(- 2\ln (\sin (r+t))-\ln \left(-\cot^2(r+t)\right)\right) \nonumber \\
B''^{(1)}_{y}&=&\frac{B'^{(0)}_{y}}{2} \ln (\cos (r+t))
\end{eqnarray}

Above equations satisfy all the Maxwell equations without a magnetic field background. Thus, taking the effect of refection and decay into account, the EM waves described by the Eqs. (\ref{28} and \ref{29}) will reduce to the EM waves obeyed by the Eqs. (\ref{ac} and \ref{ab}). We only concern the amplitude of the EM waves, so we always take the real part of the solutions in all the figures. We draw its images in Figures 4 and 5.

\section{Solution of EM waves in the region II and III }

In this section, we will study the region III ($u \le 0,~\ 0\le v \le 1 ~\ or~\ -\frac{\pi}{2}\le t-r\le \frac{\pi}{2}$ )and II ($v \le 0,~\ 0\le u \le 1 ~\ or ~\ (\frac{\pi}{2}\le t+r\le \frac{3\pi}{2}$). Similarly, repeating the processes of (\ref{E})-(\ref{H}), it gives:
\begin{eqnarray}
{\frac {\psi_{\xi}}{\psi}}-{\frac {\gamma_{\xi}}{
\gamma}}={\frac {4-4\,\sin \left( \tau+\xi \right) }{\cos \left( \tau+\xi \right) }}
\end{eqnarray}

In the region II:
\begin{eqnarray}
\label{35}
&E^{(1)}_{x}=\frac{B^{_{(0)}}_y}{4} \big(2 (r-t)-2 (r+t)+\ln (\cos (r-t))\nonumber \\ &-\ln (\cos (r+t))-\frac{4 (r-t) \sin \left(\frac{r-t}{2}\right)}{\sin \left(\frac{r-t}{2}\right)+\cos \left(\frac{r-t}{2}\right)} \nonumber\\&+\frac{4 (r+t) \sin \left(\frac{r+t}{2}\right)}{\sin \left(\frac{r+t}{2}\right)+\cos \left(\frac{r+t}{2}\right)} +2 (r-t) (\sin (r-t)-1) \sec (r-t)\nonumber\\&-2 (r-t) (\sin (r+t)-1) \sec (r+t) \nonumber\\&+\ln (\tan (r-t)+\sec (r-t)) \nonumber\\&-\ln (\tan (r+t)+\sec (r+t)) \nonumber\\&+2 \ln \left(\sin \left(\frac{r-t}{2}\right)+\cos \left(\frac{r-t}{2}\right)\right)\nonumber\\&-2 \ln \left(\sin \left(\frac{r+t}{2}\right)+\cos \left(\frac{r+t}{2}\right)\right)\big)
\end{eqnarray}

\begin{eqnarray}
\label{36}
&B^{(1)}_{y}=\frac{B^{_{(0)}}_y}{4} \big(2 (r+t)-\ln (\cos (r-t))+\ln (\cos (r+t))\nonumber \\&+\frac{4 (r-t) \sin \left(\frac{r-t}{2}\right)}{\sin \left(\frac{r-t}{2}\right)+\cos \left(\frac{r-t}{2}\right)}-\frac{4 (r+t) \sin \left(\frac{r+t}{2}\right)}{\sin \left(\frac{r+t}{2}\right)+\cos \left(\frac{r+t}{2}\right)}\nonumber \\&
-2 (r-t) (\sin (r-t)-1) \sec (r-t)\nonumber \\&+2 (r-t) (\sin (r+t)-1) \sec (r+t)\nonumber \\&-\ln (\tan (r-t)+\sec (r-t))\nonumber \\&+\ln (\tan (r+t)+\sec (r+t))\nonumber \\&+6 \ln \left(\sin \left(\frac{r-t}{2}\right)+\cos \left(\frac{r-t}{2}\right)\right)\nonumber \\&-6 \ln \left(\sin \left(\frac{r+t}{2}\right)+\cos \left(\frac{r+t}{2}\right)\right)-2 r+2 t\big)
\end{eqnarray}
 
In the region III:

\begin{eqnarray}
\label{ve}
&E^{(1)}_{x}=\big[t \tan (r-t)+t \sec (r-t)\nonumber\\&+\ln \left(\cos \left(\frac{r-t}{2}\right)-\sin \left(\frac{r-t}{2}\right)\right)-\ln \left(\cos \left(\frac{r+t}{2}\right)-\sin \left(\frac{r+t}{2}\right)\right)\big]B^{_{(0)}}_y\nonumber \\
\end{eqnarray}
\begin{eqnarray}
\label{vb}
&B^{(1)}_{y}=\big[t \tan (r-t)+t \sec (r-t)\nonumber\\&-\ln \left(\cos \left(\frac{r-t}{2}\right)-\sin \left(\frac{r-t}{2}\right)\right)\nonumber\\&+\ln \left(\cos \left(\frac{r+t}{2}\right)-\sin \left(\frac{r+t}{2}\right)\right)\big]B^{_{(0)}}_y&
\end{eqnarray}
   \begin{figure}
  \centering
  \includegraphics[width=.3\textwidth]{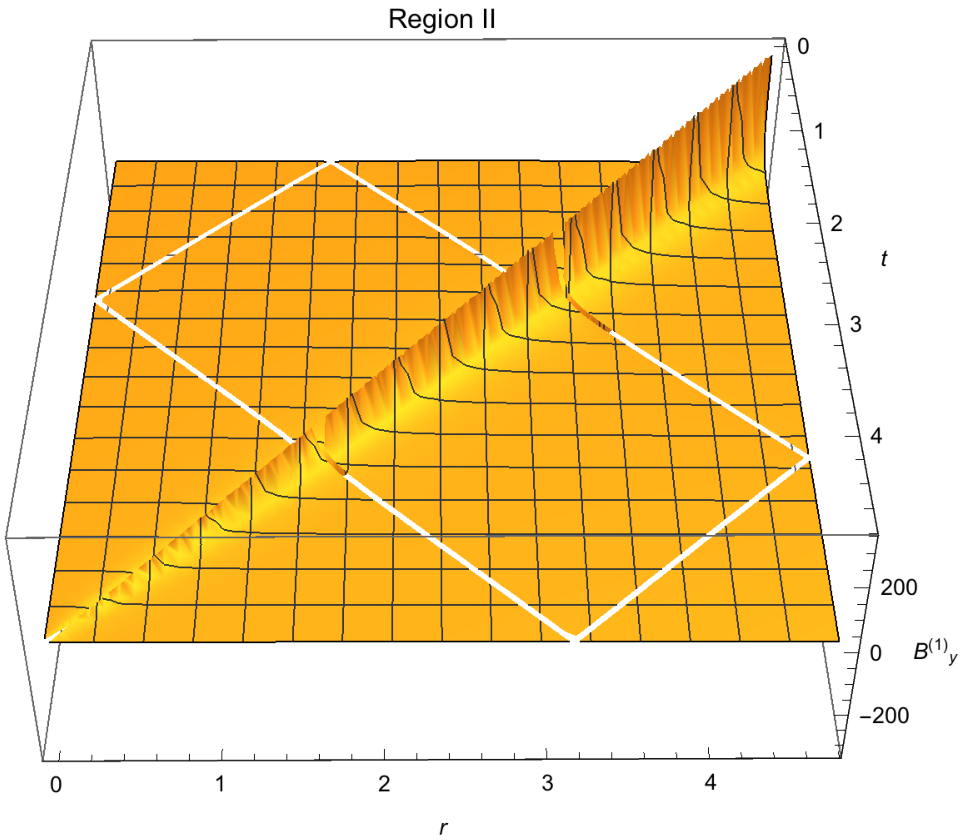} 
    \includegraphics[width=.3\textwidth]{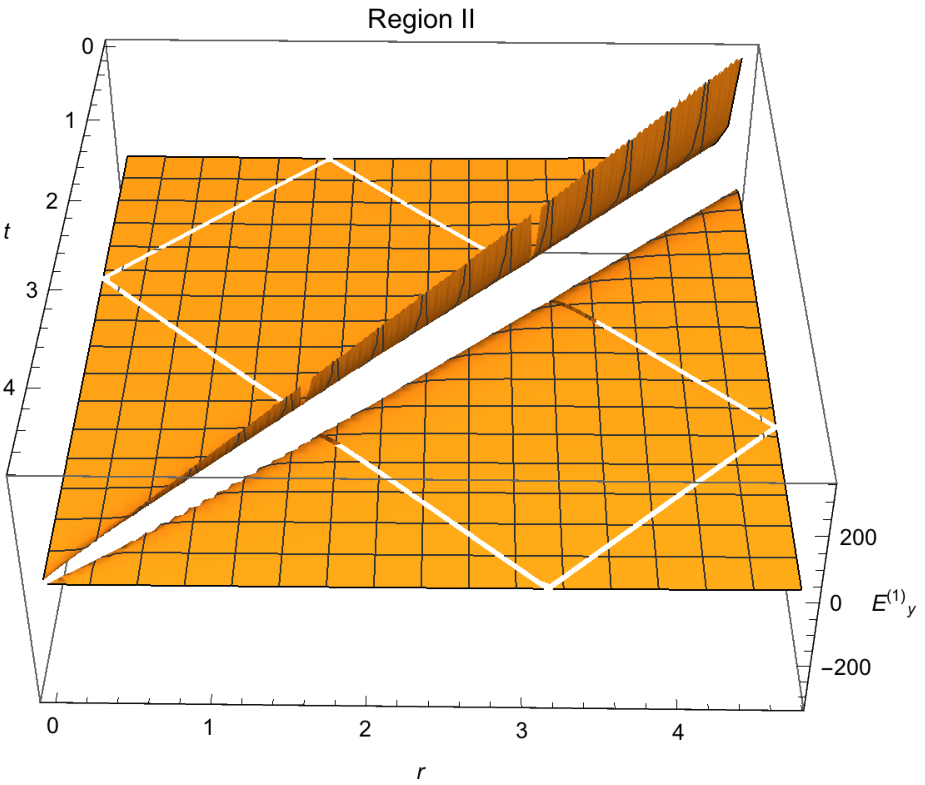} 
        \includegraphics[width=.3\textwidth]{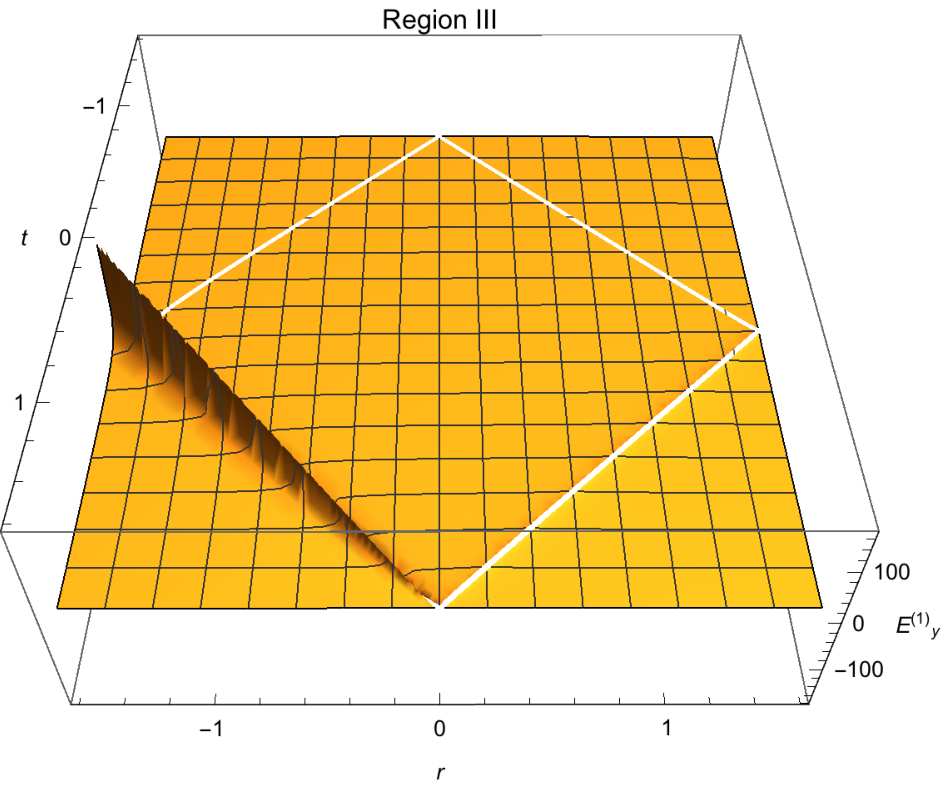} 
                \includegraphics[width=.3\textwidth]{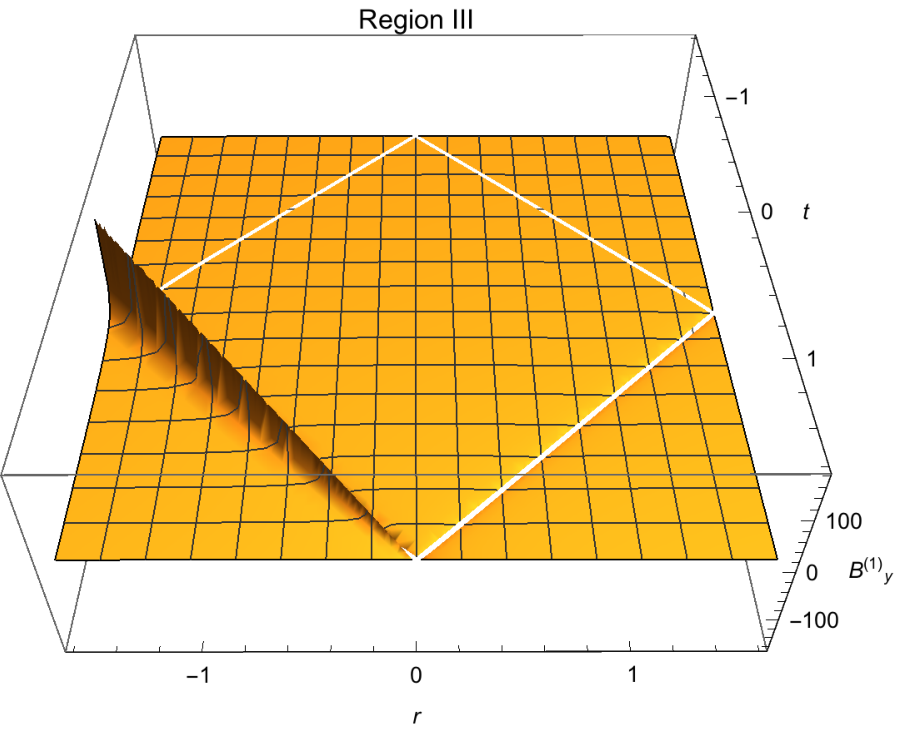} 
    \caption{The first two diagrams describe the Eqs. (\ref{35} and \ref{36}) in 3D plot, and the last two diagrams describe the Eqs. (\ref{ve} and \ref{vb}) in 3D plot.  }
\end{figure}

As you can see in Fig.7, EM waves have suffered a singularity at the $t+r=\frac{3}{2}\pi$ and $t-r=\frac{\pi}{2}$ respectively. It is very easy to understand because the PGWs have singularity at the $t+r=\frac{3}{2}\pi$ and  $t-r=\frac{\pi}{2}$ \cite{Ferrari1987On}.
The EM waves perturbed by PGW  also have an accumulation effect (see Figs. 8 and 9) which is similar to the result of H. Wen\cite{Wen:2014wxa}. The region II and III interact with the magnetic field background also is finite. When the PGWs totally cross the region the magnetic field, the Eqs. (\ref{EE} and \ref{BB}) have also become:

\begin{equation}
\label{EE1}
{\frac {\partial ^{2}}{\partial {t}^{2}}}{\it E^{(1)}_{x}} \left( t,r
 \right)-{\frac {\partial ^{2}}{\partial {r}^{2}}}{\it E^{(1)}_{x}} \left( t,r \right)  =0
\end{equation}
\begin{equation}
\label{BB1}
{\frac {\partial ^{2}}{\partial {t}^{2}}}{\it B^{(1)}_{y}} \left( t,r
 \right)-{\frac {\partial ^{2}}{\partial {r}^{2}}}{\it B^{(1)}_{y}} \left( t,r \right)  =0
\end{equation}
~\\
\begin{figure}
  \centering
  \includegraphics[width=.3\textwidth]{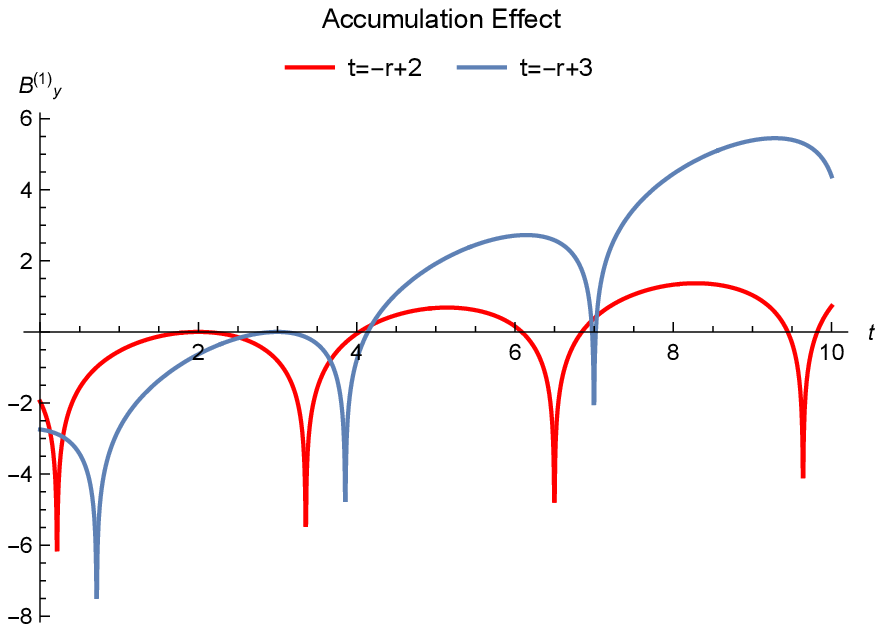} 
    \includegraphics[width=.3\textwidth]{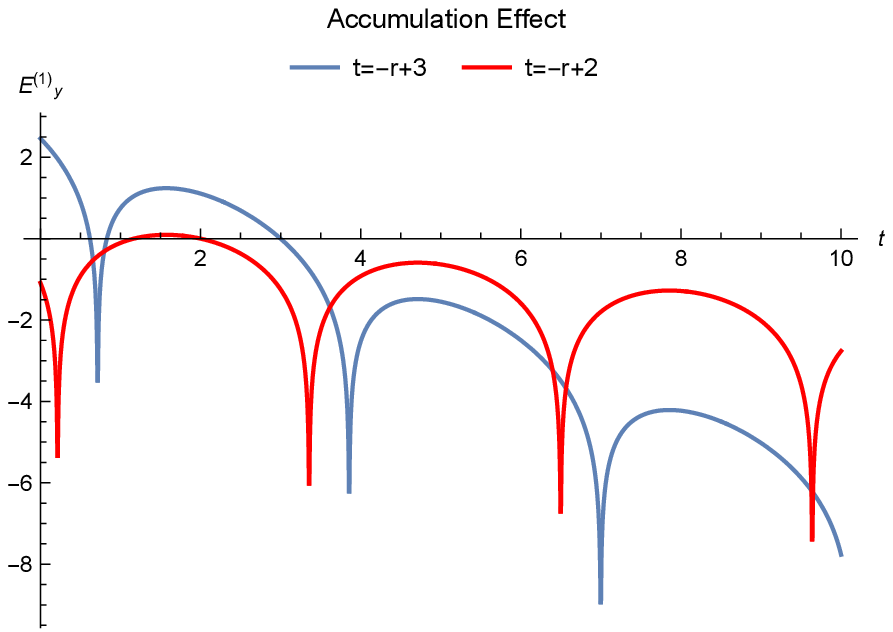} 
    \caption{In region II, the accumulation effect is showed.  }
\end{figure}
    \begin{figure}
  \centering
  \includegraphics[width=.3\textwidth]{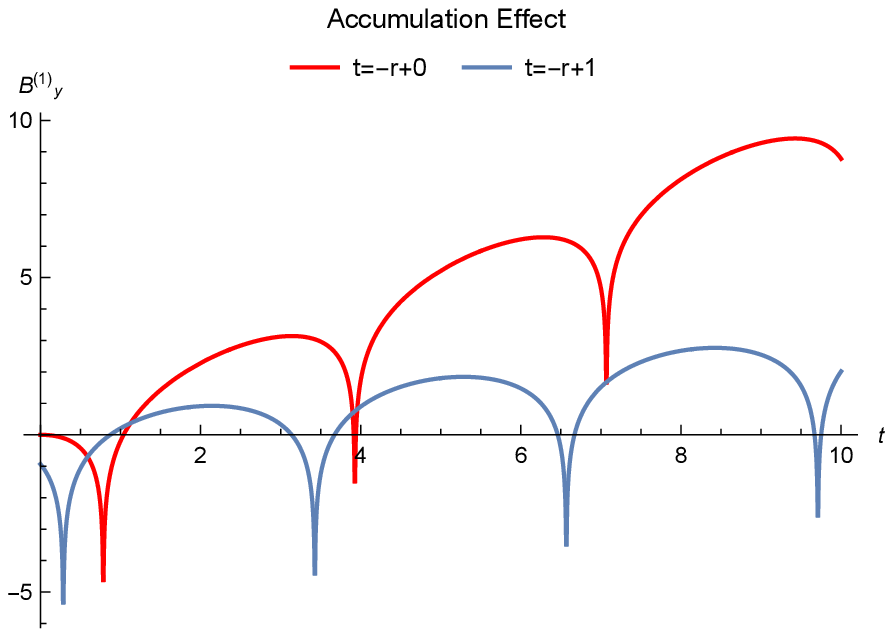} 
    \includegraphics[width=.3\textwidth]{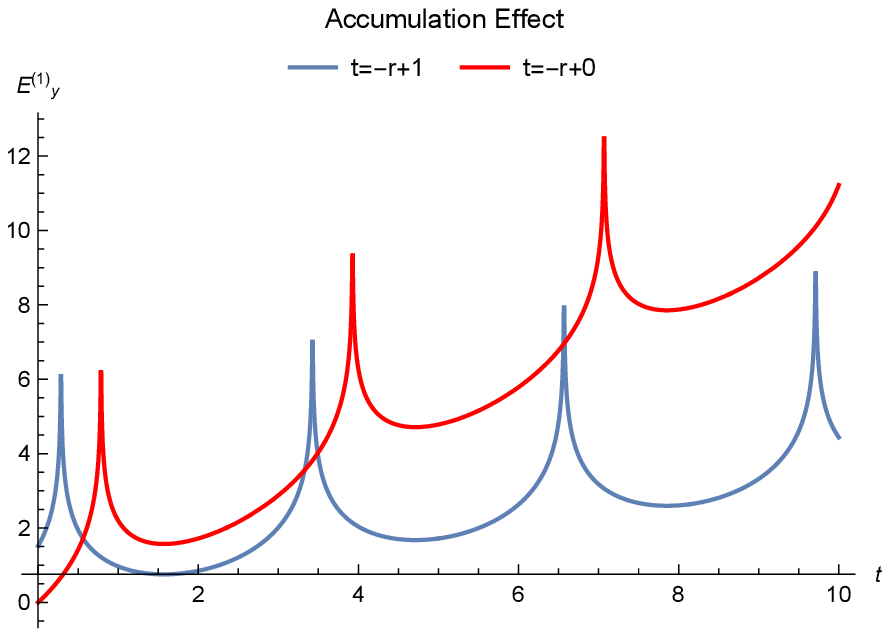} 
    \caption{In region III, the accumulation effect is showed. }
\end{figure}

 Therefore, the solutions of EM waves with respect to the Eqs. (\ref{35},\ref{36} and \ref{ve},\ref{vb}) propagate in spacetime without a magnetic field, which must suffer a reflection and decay(caused by the different magnetic field background). Noticing the first and second term of Eqs. (\ref{ve}) and (\ref{vb}), it causes the accumulation effect which should be disappear in the metric (\ref{9}) without the magnetic field. And considering the equations:
    \begin{eqnarray}
\label{ve1}
E'^{(1)}_{x}=[-\ln \left(\cos \left(\frac{r+t}{2}\right)-\sin \left(\frac{r+t}{2}\right)\right) ~\nonumber \\+C(r)]B^{_{(0)}}_y
\end{eqnarray}
\begin{eqnarray}
\label{vb1}
B'^{(1)}_{y}=[\ln \left(\cos \left(\frac{r+t}{2}\right)-\sin \left(\frac{r+t}{2}\right)\right) ~\nonumber \\+C(r)]B^{_{(0)}}_y
\end{eqnarray}

where C(r) is a constant which has a relationship with the accumulation effect.

 Eqs.(\ref{ve1} and \ref{vb1}) satisfy Eqs.(\ref{EE1} and \ref{BB1}) respectively and also satisfy all the Maxwell equations(\ref{eq2}) where $B^{(0)}_{y}=0$. The term of remanent in Eqs. (\ref{ve}) and (\ref{vb}) should regard as reflection or decay. 

Hence, observer at a long distant will receive a PGW and EM waves portrayed by the Eqs. (\ref{ve1} and \ref{vb1}) (see Fig.10).
 \begin{figure}
  \centering
    \includegraphics[width=.3\textwidth]{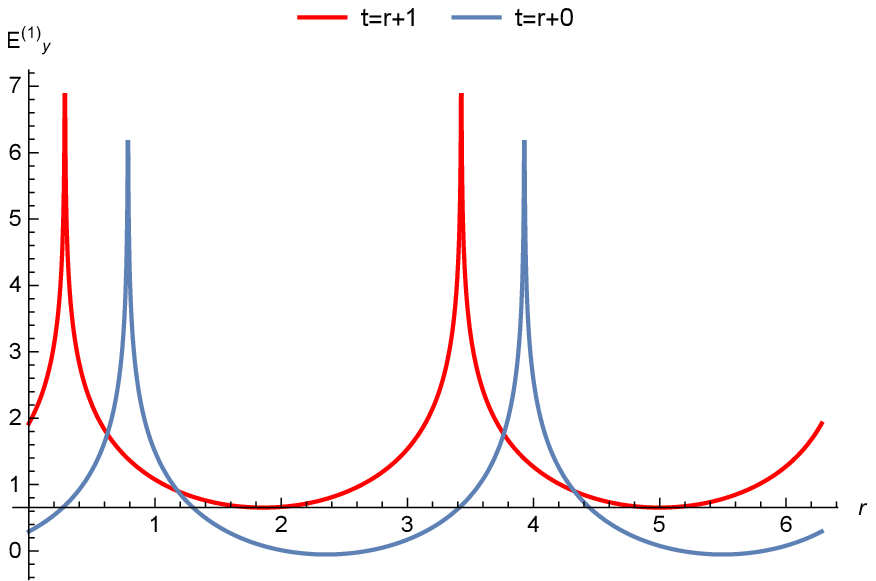} 
     \includegraphics[width=.3\textwidth]{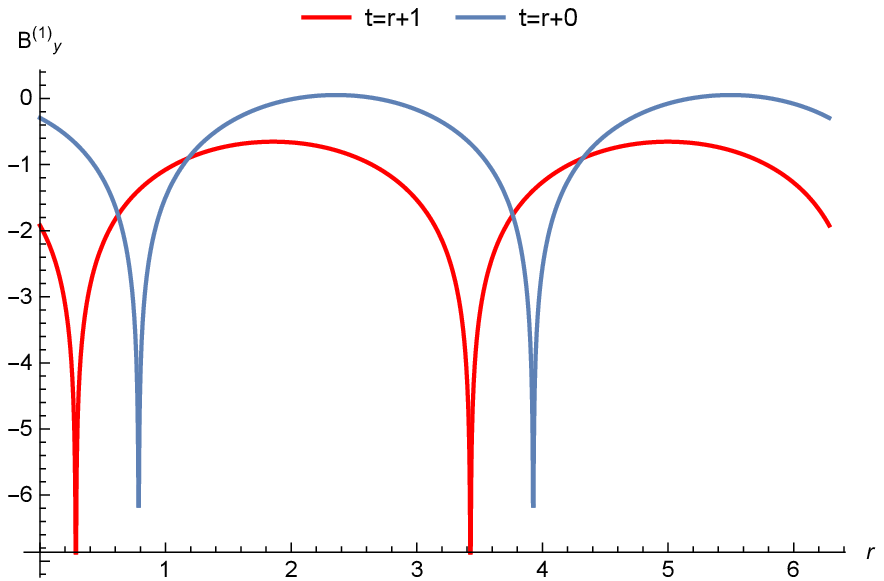} 
    \caption{ These figures describe the Eqs. (\ref{ve1} and \ref{vb1}) and we assume that the EM waves through out the magnetic field at $t=1$ }
\end{figure}
\begin{figure}
  \centering
 \includegraphics[width=.3\textwidth]{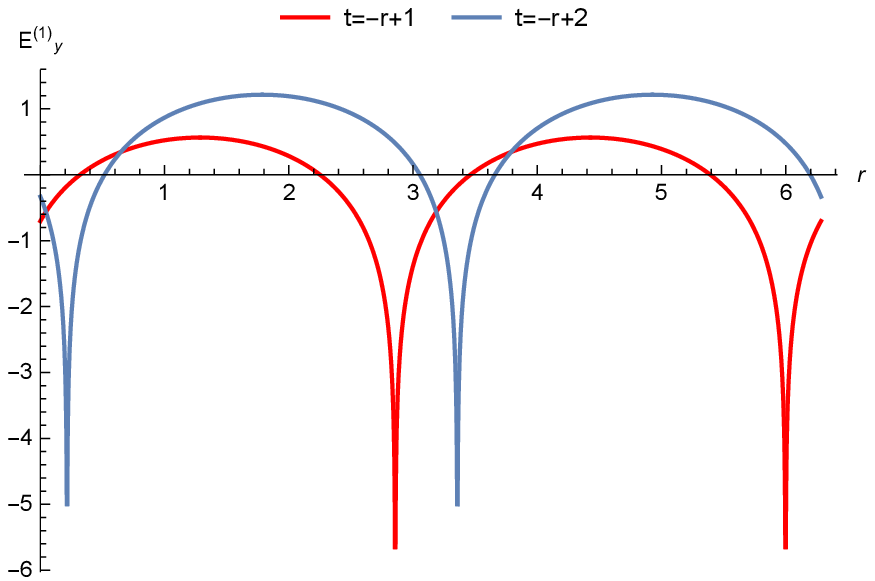} 
 \includegraphics[width=.3\textwidth]{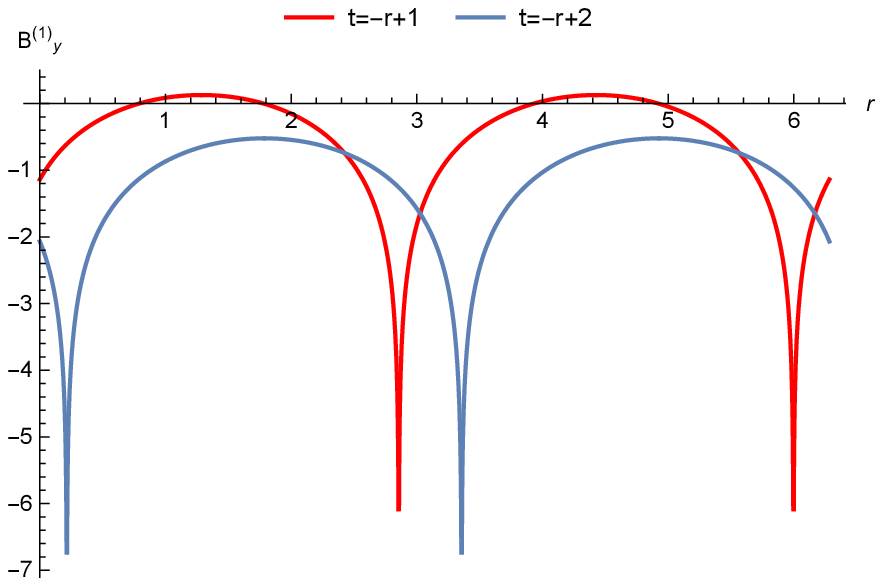} 
    \caption{These figures describe the Eqs. (\ref{ue1} and \ref{ub1}) and we also assume that the EM waves through out the magnetic field at $t=1$ }
\end{figure}

We have repeated the same process to the region  II, and we also have:
 \begin{eqnarray}
 \label{ue1}
&E'^{(1)}_{x}=\frac{B^{_{(0)}}_y}{4} (\ln (\cos (r-t))+\ln (\tan (r-t)+\sec (r-t))\nonumber\\&+2 \ln \left(\sin \left(\frac{r-t}{2}\right)+\cos \left(\frac{r-t}{2}\right)\right)+C(r))
\end{eqnarray}
\begin{eqnarray}
\label{ub1}
&B'^{(1)}_{y}=\frac{B^{_{(0)}}_y}{4} (-\ln (\cos (r-t))-\ln (\tan (r-t)+\sec (r-t))\nonumber\\&+6 \ln \left(\sin \left(\frac{r-t}{2}\right)+\cos \left(\frac{r-t}{2}\right)\right)+C(r))
\end{eqnarray}
Fig. 11 sketches the EM waves in the metric (\ref{8}) without the magnetic field background. 
\section{Summary And Conclusion}
As you can see in the Figs 12, a surface $\Delta s$ in region III which can receive electric field or magnetic field of the EM waves, can be judged the PWs whether or not to collide. And comparing the Figs.4 and 5 with Figs.10 and 11, we find that their amplitudes and phases are different( Specially, the maximum value of amplitude in Eqs. (32, 33) are only about half of that in Eqs. (41-44) , but the waveforms are somewhat similar. If the magnetic field is strong enough, we will accurately test whether the colliding region of PGWs really happened. On the other hand, if the region of the magnetic field is large enough, we will also precisely test the general relativity by observing the accumulation effect in region II and III, because other theories will absolutely give a different result.
  
 \begin{figure}
  \centering
    \includegraphics[width=.3\textwidth]{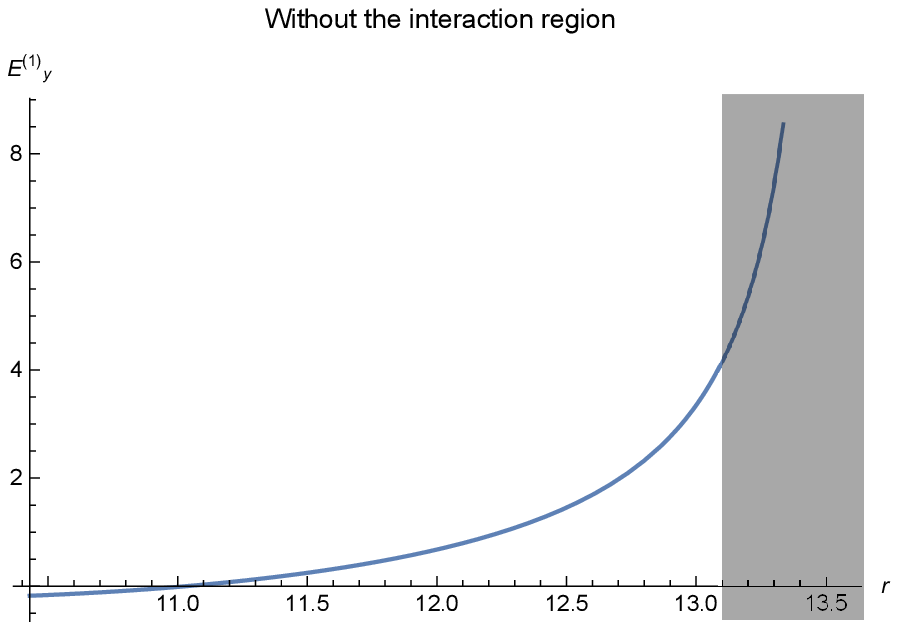} 
     \includegraphics[width=.3\textwidth]{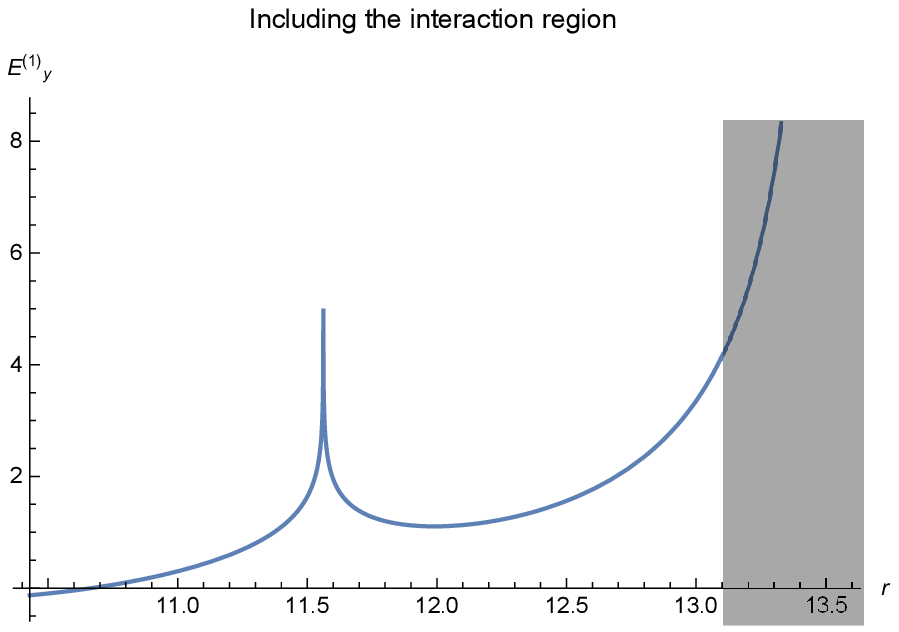} 
 \includegraphics[width=.3\textwidth]{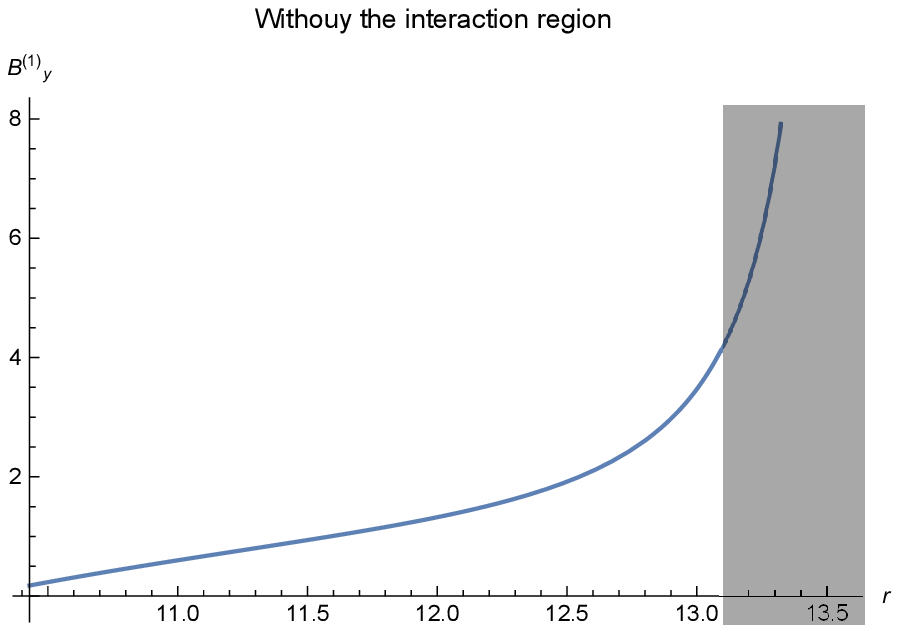} 
 \includegraphics[width=.3\textwidth]{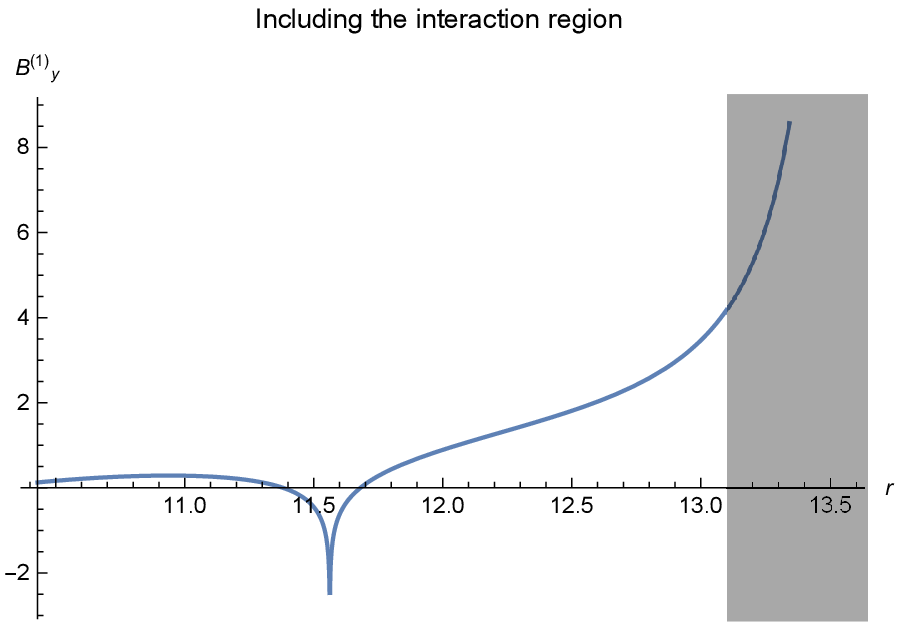} 
    \caption{The gray area is described a region which the method we used cannot accuratly predict the EM waves, becausing the higher orders cannot be omitted when encountering curvature singularity.  }
\end{figure}

In present work, we have exactly calculated out the solutions of EM field perturbed by CPWs. Firstly, in the region II and III, EM waves created by the perturbed spacetime have propagated in the same direction as gravitational waves. And it exists an accumulation effect and a singularity at the curvature singularity of the spacetime in the PGWs. Moreover, we also accurately work out the solution of EM waves after suffering a reflection and decay(caused by the different magnetic field background). Secondly, in the colliding region, the solution of EM waves produced by the perturbed colliding region has precisely worked out. These EM waves have spread along u and v directions, but it does not have an accumulation effect which different to before. And we also exactly figure out the solution of these EM waves after crossing to the region without magnetic background.

At last, there are still a lot of interesting works to be done in the field of the CPWs.

\bibliographystyle{unsrt}   
\bibliography{citedpapers}   

\end{document}